\documentclass[11pt]{article}

\usepackage{graphicx,amsmath,amsfonts,amssymb, dcolumn,amsthm}

\begin{document}

\title{\textbf{de Sitter gauge theories and induced gravities}}
\author{\textbf{R.~F.~Sobreiro$^1$}\thanks{sobreiro@if.uff.br}\ ,  \textbf{A.~A.~Tomaz$^1$}\thanks{tomaz@if.uff.br}\ ,\textbf{V.~J.~Vasquez Otoya$^2$}\thanks{victor.vasquez@ifsudestemg.edu.br}\\\\
\textit{{\small $^1$UFF $-$ Universidade Federal Fluminense,}}\\
\textit{{\small Instituto de F\'{\i}sica, Campus da Praia Vermelha,}}\\
\textit{{\small Avenida General Milton Tavares de Souza s/n, 24210-346,}}\\
\textit{{\small Niter\'oi, RJ, Brasil.}}\and
\textit{{\small $^2$IFSEMG $-$ Instituto Federal de Educa\c{c}\~ao, Ci\^encia e Tecnologia,}} \\
\textit{{\small Rua Bernardo Mascarenhas 1283, 36080-001,}}\\
\textit{{\small Juiz de Fora, MG, Brasil}}}
\date{}
\maketitle

\begin{abstract}
Pure de Sitter, anti de Sitter, and orthogonal gauge theories in four-dimensional Euclidean spacetime are studied. It is shown that, if the theory is asymptotically free and a dynamical mass is generated, then an effective geometry may be induced and a gravity theory emerges. The asymptotic freedom and the running of the mass might account for an In\"on\"u-Wigner contraction which induces a breaking of the gauge group to the Lorentz group, while the mass itself is responsible for the coset sector of the gauge field to be identified with the effective vierbein. Furthermore, the resulting local isometries are Lorentzian for the anti de Sitter group and Euclidean for the de Sitter and orthogonal groups.
\end{abstract}

\section{Introduction}\label{intro}

The fact that three of the four fundamental interactions are gauge theories provides one of the main motivations for the construction of gauge theories of gravity \cite{Utiyama:1956sy,Kibble:1961ba,Sciama:1964wt}. Therefore, two fields are introduced, the vierbein $e$ and the spin connection $\omega$. Specific composite fields constructed from them provide the geometric properties of spacetime \cite{Mardones:1990qc,Zanelli:2005sa}. It turns out that the deep relation between the fields of gravity and spacetime spoils the possibility of a quantum description of gravity independent of the background geometry, \emph{i.e.}, a quantum field should not actually depend on parameters that also fluctuate. Moreover, even in a background dependent quantization, the Einstein-Hilbert action itself is not enough to ensure perturbative quantum stability of gravity \cite{'tHooft:1974bx,Deser:1974cz,Deser:1974cy}.

To circumvent these problems, many other theories have been proposed, still applying the gauge theoretical approach, by generalizing the gauge groups and their respective actions. In particular, it is worth mentioning the de Sitter groups $SO(m,n)$, with $(m+n)=5$, in four-dimensional spacetime \cite{Stelle:1979va,Stelle:1979aj,Tseytlin:1981nu}, need a gauge symmetry breaking; in this way the vierbein can emerge. The most used method for the breaking is the Higgs mechanism, which needs an extra set of scalar fields, see for instance \cite{Stelle:1979va,Stelle:1979aj,Tseytlin:1981nu,Hehl:1976my,MacDowell:1977jt,Pagels:1983pq,Mahato:2004zi,Tresguerres:2008jf,Mielke:2010zz}. In these works, together with de Sitter groups, several groups are considered as well as different starting actions that encodes gravity as a limit. This last feature is the essence of the so called emergent gravities, see also \cite{Sindoni:2011ej,Kolekar:2011gw} and references therein.

The present work is about de Sitter gauge theories in four-dimensional Euclidean spacetime, where the ideas developed in \cite{Sobreiro:2007pn,Sobreiro:2010ji,Sobreiro:2010qf} are further exploited. The starting action is the massless pure Yang-Mills action in a four-dimensional spacetime. Thus, it is renormalizable at least to all orders in perturbation theory \cite{Piguet:1995er}. The choice of an Euclidean space is not accidental, it follows from the fact that any quantum field theory is actually treatable only in Euclidean spaces (even perturbatively, where a Wick rotation is needed for reliable quantum computations). Moreover, in an Euclidean manifold, space and time are indistinguishable, and thus time evolution of any physical system becomes, at least, unclear. On the other hand, non-Abelian gauge theories have two main effects. First, the theory is perturbatively asymptotically free \cite{Gross:1973id,Politzer:1973fx}. Second, dynamical mass parameters might emerge at non-perturbative level as the coupling parameter increases, see for instance \cite{Gribov:1977wm,Zwanziger:1992qr,Sobreiro:2005ec,Dudal:2005na,Aguilar:2010ze,Dudal:2011gd} and references therein.

Then, we use both effects to show that an induced gravity theory can emerge naturally, where the running parameters induce an In\"on\"u-Wigner contraction \cite{Inonu:1953sp} from de Sitter algebra to Poincar\'e algebra. Since the Poincar\'e group is not a symmetry of the original action, the symmetry is actually broken to a Lorentz type group $SO(m!-1,n)$ and an effective gravity emerges. Under that approach, Newton and cosmological constants are associated with the mass and coupling parameters. The local isometries of the deformed spacetime depend on the value $m$, and for $m=2$ this results in a Lorentz local symmetry. This last result can be interpreted as the rising of the equivalence principle. We also provide a formal analysis for the induced gravity by considering the corresponding fiber bundle theory \cite{Daniel:1979ez,Trautman:1981fd,Nakahara:1990th}.

In this work, Section 2 is devoted to the foundations of de Sitter gauge theories in four-dimensional Euclidean spacetime. In Section 3 the In\"on\"u-Wigner contraction is performed and the symmetry breaking is shown. The induced gravity is discussed in Section 4. The mathematical analysis of the model is displayed in Section 5. Finally, our conclusions and final remarks are found in Section 6.

\section{de Sitter gauge theories in four Euclidean dimensions}

We consider a gauge theory based on the group $SO(m,n)$ with $m+n=5$ and $m\in\{0,1,2\}$ while the spacetime is an Euclidean four-dimensional differential manifold $\mathbb{R}^4$. The gauge group is then the orthogonal group for $m=0$, the de Sitter group for $m=1$ and anti de Sitter for $m=2$. Except when necessary, we shall indistinguishably call the generic group, with arbitrary $m$, by de Sitter group. The algebra of the group is given by
\begin{equation}
\left[J^{AB},J^{CD}\right]=-\frac{1}{2}\left[\left(\eta^{AC}J^{BD}+\eta^{BD}J^{AC}\right)-\left(\eta^{AD}J^{BC}+\eta^{BC}J^{AD}\right)\right]\;,\label{alg1}
\end{equation}
where $J^{AB}$ are the $10$ anti-hermitian generators of the gauge group, antisymmetric in their indices. Caption Latin indices are chosen to run as $\{5,0,1,2,3\}$. The $SO(m,n)$ group defines a five-dimensional flat space, $\mathbb{R}^{m,n}_S$, with invariant Killing metric given by $\eta^{AB}\equiv\mathrm{diag}(\epsilon,\varepsilon,1,1,1)$ with $\epsilon=(-1)^{(2-m)!}$ and $\varepsilon=(-1)^{m!+1}$. We stress out that this is a gauge theory, pure and simple, it has no relation with the spacetime dynamics.

The de Sitter group may be decomposed as a direct product, $SO(m,n)\equiv SO(m!-1,n)\otimes S(4)$ where $S(4)\equiv SO(m,n)/SO(m!-1,n)$ is a symmetric coset space with four degrees of freedom. This decomposition is carried out by projecting the group space in the fifth coordinate $A=5$. Defining then $J^{5a}=J^a$, where small Latin indices run as $\{0,1,2,3\}$, the algebra \eqref{alg1} decomposes as
\begin{eqnarray}
\left[J^{ab},J^{cd}\right]&=&-\frac{1}{2}\left[\left(\eta^{ac}J^{bd}+\eta^{bd}J^{ac}\right)-\left(\eta^{ad}J^{bc}+\eta^{bc}J^{ad}\right)\right]\;,\nonumber\\
\left[J^a,J^b\right]&=&-\frac{\epsilon}{2}J^{ab}\;,\nonumber\\
\left[J^{ab},J^c\right]&=&\frac{1}{2}\left(\eta^{ac}J^b-\eta^{bc}J^a\right)\;,\label{alg2}
\end{eqnarray}
where $\eta^{ab}\equiv\mathrm{diag}(\varepsilon,1,1,1)$.

The fundamental field is the 1-form gauge connection, an algebra-valued quantity in the adjoint representation $Y=Y^A_{\phantom{A}B}J_A^{\phantom{A}B}=A^a_{\phantom{a}b}J_a^{\phantom{a}b}+\theta^aJ_a$, whose gauge transformation is
\begin{equation}
Y\longmapsto g^{-1}\left(\frac{1}{\kappa}\mathrm{d}+Y\right)g\;,\Big|\;g\in SO(m,n)\;,\label{gt0}
\end{equation} 
where, obviously, $\kappa$ is a dimensionless coupling parameter and $\mathrm{d}$ the exterior derivative. At infinitesimal level, we have $Y\longmapsto Y+\nabla\zeta$,  where $g=\exp{(\kappa\zeta)}\approx1+\kappa\zeta$ and $\nabla=\mathrm{d}+\kappa Y$ is the full covariant derivative. This transformation decomposes as
\begin{eqnarray}
 {A}^a_{\phantom{a}b}&\longmapsto& {A}^a_{\phantom{a}b}+\mathrm{D}\alpha^a_{\phantom{a}b}-\frac{\epsilon\kappa}{4}\left(\theta^a\xi_b-\theta_b\xi^a\right)\;,\nonumber\\
\theta^a&\longmapsto&\theta^a+\mathrm{D}\xi^a+\kappa\alpha^a_{\phantom{a}b}\theta^b\;.\label{gt2}
\end{eqnarray}
where $\zeta=\alpha^a_{\phantom{a}b}J_a^{\phantom{a}b}+\xi^aJ_a$ and $\mathrm{D}=\mathrm{d}+\kappa A$ is the covariant derivative with respect to the sector $SO(m!-1,n)$.

The operator $\nabla^2$ defines the 2-form field strength, $F=\mathrm{d}Y+\kappa YY$, which decomposes as $F=\left(\Omega^a_{\phantom{a}b}-\frac{\epsilon\kappa}{4}\theta^a\theta_b\right)J_a^{\phantom{a}b}+K^aJ_a$ where $\Omega^a_{\phantom{a}b}=\mathrm{d}A^a_{\phantom{a}b}+\kappa A^a_{\phantom{a}c}A^c_{\phantom{c}b}$ and\footnote{It might be evident for the reader that the $K^a$ sector will be identified with torsion and, in fact, this is the intention (c.f. Section \ref{induced}). Nevertheless, the minus sign seems to be not correct from standard conventions. The explanation is that in \eqref{alg1} a global minus sign appear in order to adjust our conventions of anti-hermitian generators. As a consequence, the last of \eqref{alg2} also appear with a different global sign and this is the relevant commutation relation for $K^a$.} $K^a=\mathrm{D}\theta^a=\mathrm{d}\theta^a-\kappa A^a_{\phantom{b}b}\theta^b$.

To construct the most general gauge invariant action we demand: i.) Absence of mass parameters in the starting theory; any mass parameter must appear from dynamical effects. ii.) Parity symmetry in both spaces, namely the $\mathbb{R}^4$ spacetime and $\mathbb{R}^{m,n}_S$ gauge space. iii.) Locality and renormalizability. It turns out that the gauge invariant action fulfilling these requirements is the usual Yang-Mills action
\begin{eqnarray}
S_{\mathrm{YM}}&=&\frac{1}{2}\int F^A_{\phantom{A}B}{*} F_A^{\phantom{A}B}\nonumber\\
&=&\frac{1}{2}\int\left[\Omega^a_{\phantom{a}b}{*}\Omega_a^{\phantom{a}b}+\frac{1}{2}K^a{*} K_a-\frac{\epsilon\kappa}{2}\Omega^a_{\phantom{a}b}{*}(\theta_a\theta^b)+\frac{\kappa^2}{16}\theta^a\theta_b{*}(\theta_a\theta^b)\right]\;,\label{ym0}
\end{eqnarray}
where ${*}$ denotes the Hodge dual operation in spacetime.

A few observations are in order: i.) An important feature of the present action is the absence of mass parameters. Usually, in de Sitter gravity \cite{Stelle:1979va,Stelle:1979aj,Tseytlin:1981nu}, the field $\theta^a$ possesses components $\theta^a_\mu$ that carry dimension 0 and always appear with a mass scale factor (the cosmological constant) to adjust the correct UV dimension of a connection component. In the present model the components $\theta^a_\mu$ carry UV dimension 1 and then cannot be directly associated with the coframes. ii.) To quantize the model, a gauge fixing is needed. From the very beginning, the action is the usual Yang-Mills action for a semi-simple Lie group; then, it is renormalizable, at least to all orders in perturbation theory, depending on the gauge choice \cite{Piguet:1995er}. The main difference between the present action and the $SU(N)$ Yang-Mills theories lies on the fact that $SO(m,n)$ may be non-compact, a property that might spoil the unitarity of the theory. We shall discuss this in more detail in Section 5. iii.) As a non-Abelian theory, it is asymptotically free \cite{Gross:1973id,Politzer:1973fx}. As a consequence, a non-pertubative behavior is expected at the infrared regime, which becomes more evident by means of an increasing of the coupling parameter $\kappa$. iv.) The non-linearity of the theory also favors the condensation of composite operators and thus the possibility of dynamical mass parameters to emerge \cite{Dudal:2005na,Aguilar:2010ze}. On the other hand, at least one mass parameter is required for quantization improvements in order to fix the so called Gribov ambiguities \cite{Gribov:1977wm,Zwanziger:1992qr,Sobreiro:2005ec,Dudal:2011gd}.

\section{In\"on\"u-Wigner contraction}

Independently of the physical mechanism, a mass scale is assumed and is denoted here by $\gamma$. The existence of a mass allows a rescaling of the fields. In particular, to make contact with gravity, we employ the rescaling
\begin{eqnarray}
A&\longmapsto&\kappa^{-1}A\;,\nonumber\\
\theta&\longmapsto&\kappa^{-1}\gamma\theta\;.\label{resc1}
\end{eqnarray}
See, for instance, \cite{Stelle:1979va}, where a similar rescaling is used. The transformations \eqref{resc1} are not accidental. Both sectors are rescaled with $\kappa^{-1}$ in order to factor out the coupling parameter outside the action, a standard procedure in Yang-Mills theories \cite{Itzykson:1980rh}. On the other hand, the mass parameter affects only the $\theta$-sector, transforming it in a field with dimensionless components. It turns out that this is the unique possibility if one wishes to identify $\theta$ with a vierbein field. If also $A$ is rescaled with a mass factor, then it would never be possible to identify it with the spin connection. The action \eqref{ym0} is then rescaled to
\begin{equation}
S=\frac{1}{2\kappa^2}\int\left[\overline{\Omega}^a_{\phantom{a}b}{*}\overline{\Omega}_a^{\phantom{a}b}+\frac{\gamma^2}{2}\overline{K}^a{*}\overline{K}_a-\frac{\epsilon\gamma^2}{2}\overline{\Omega}^a_{\phantom{a}b}{*}(\theta_a\theta^b)+\frac{\gamma^4}{16}\theta^a\theta_b{*}(\theta_a\theta^b)\right]\;,\label{ym1}
\end{equation}
where $\overline{\Omega}^a_{\phantom{a}b}=\mathrm{d} {A}^a_{\phantom{a}b}+ {A}^a_{\phantom{a}c} {A}^c_{\phantom{c}b}$, $\overline{K}^a=\mathrm{D}\theta^a$ and the covariant derivative is now $\mathrm{D}=\mathrm{d}+A$. Moreover a reparameterization of the $SO(m,n)$ generators is required due to the existence of a mass scale, \emph{i.e.}, a stereographic projection is now allowed if one identifies the mass parameter with the radius of the gauge manifold $\mathbb{R}^{m,n}_S$, \emph{i.e.}, $J^a=-\kappa\gamma^{-1}P^a+\kappa^{-1}\gamma \overline{x}^a\overline{x}_bP^b$, where $\overline{x}^a$ are stereographic coordinates in group subspace $\mathbb{R}_S^{(m!-1),n}$. It follows that $\theta\longmapsto\kappa^{-1}\gamma\theta=-\theta^aP_a+\kappa^{-2}\gamma^2\theta^a\overline{x}_a\overline{x}_bP^b$. Thus, the algebra of the de Sitter group \eqref{alg1} is now given by
\begin{eqnarray}
\left[J^{ab},J^{cd}\right]&=&-\frac{1}{2}\left[\left(\eta^{ac}J^{bd}+\eta^{bd}J^{ac}\right)-\left(\eta^{ac}J^{bc}+\eta^{bc}J^{ad}\right)\right]\;,\nonumber\\
\left[J^a,J^b\right]&=&-\frac{\epsilon\gamma^2}{2\kappa^2}J^{ab}\;,\nonumber\\
\left[J^{ab},J^c\right]&=&\frac{1}{2}\left(\eta^{ac}J^b-\eta^{bc}J^a\right)\;.\label{alg3}
\end{eqnarray}

The presence of a mass parameter has been used and we now explore the asymptotic freedom of the model. The fact that, at low energies, the coupling parameter $\kappa$ increases enforces that the quantity $\gamma^2/\kappa^2$ is very small for some non-perturbative scales. This property implies that the algebra of Eq.~\eqref{alg3} suffers an In\"on\"u-Wigner contraction \cite{Inonu:1953sp}, contracting down to the Poincar\'e algebra, \emph{i.e.}, the second commutator of \eqref{alg3} is replaced by $\left[P^a,P^b\right]=0$, where the projected generator turns out to reduce to the usual translational one $J^a\longmapsto -\kappa\gamma^{-1}P^a$ and thus $\theta\longmapsto-\theta^aP_a$. The gauge symmetry is then dynamically deformed to the Poincar\'e group, $SO(m,n)\longrightarrow ISO(m!-1,n)$, for some values $\kappa$ in the strong coupling regime.

The In\"on\"u-Wigner contraction induces a symmetry breaking of the action \eqref{ym0} which is invariant under $SO(m,n)$ but not under $ISO(m!-1,n)$ because $ISO(m!-1,n)\nsubseteq SO(m,n)$. On the other hand, the group $SO(m!-1,n)$ is a subgroup of both groups,  $ISO(m!-1,n)\supset SO(m!-1,n)\subset SO(m,n)$. Thus, the In\"on\"u-Wigner contraction actually implies on a symmetry breaking, $SO(m,n)\longrightarrow SO(m!-1,n)$. Under the $SO(m!-1,n)$ gauge symmetry, the transformations \eqref{gt2} reduce to
\begin{eqnarray}
 A^a_{\phantom{a}b}&\longmapsto& {A}^a_{\phantom{a}b}+\mathrm{D}\alpha^a_{\phantom{a}b}\;,\nonumber\\
\theta^a&\longmapsto&\theta^a-\alpha^a_{\phantom{a}b}\theta^b\;.\label{gt3}
\end{eqnarray}
where \eqref{resc1} was assumed.

\section{Induced gravity}\label{induced}

In \cite{Obukhov:1998gx}, it was shown that a three-dimensional spacetime can be deformed in several ways by an $SU(2)$ gauge theory. It turns out that this idea can be generalized to four dimensions in the present theory: Every configuration $(A,\theta)$ defines an effective geometry $(\omega,e)$ by means of a mapping from each point $x\in\mathbb{R}^4$ to a point $X\in\mathbb{M}^4$ of the deformed space. In order to preserve the algebraic structure already defined in $\mathbb{R}^4$, it is demanded that this mapping is an isomorphism \cite{Nakahara:1990th}. The local gauge group $SO(m!-1,n)$ defines, at each point of the mapping, the isometries of the tangent space $T_X(\mathbb{M})$. Thus, $\theta$ and $A$ can be identified with the vierbein $e$ and spin connection $\omega$, respectively,
\begin{eqnarray}
\omega^{\mathfrak{ab}}_\mu(X)dX^\mu&=&\delta^{\mathfrak{a}}_a\delta^{\mathfrak{b}}_bA^{ab}_\mu(x)dx^\mu\;,\nonumber\\
e^{\mathfrak{a}}_\mu(X)dX^\mu&=&\delta_a^{\frak{a}}\theta^a_\mu(x)dx^\mu\;.\label{id2}
\end{eqnarray}
In expressions \eqref{id2}, latin indices $\mathfrak{a}$, $\mathfrak{b}$... belong to the tangent space $T_X(\mathbb{M})$. For the Hodge duals, we impose the simplest mapping
\begin{equation}
\star f(X)=\ast f(x)\;,\label{id4}
\end{equation}
where $f$ is a generic $p$-form and $\star$ is the Hodge dual in $\mathbb{M}^4$. The mapping provides then
\begin{equation}
S=\frac{\gamma^2}{4\kappa^2}\int\left[\frac{2}{\gamma^2}R^\mathfrak{a}_{\phantom{a}\mathfrak{b}}\star R_\mathfrak{a}^{\phantom{a}\mathfrak{b}}+T^\mathfrak{a}\star T_\mathfrak{a}-\frac{\epsilon}{2}\epsilon_\mathfrak{abcd}R^\mathfrak{ab}e^\mathfrak{c}e^\mathfrak{d}+\frac{\gamma^2}{16}\epsilon_\mathfrak{abcd}e^\mathfrak{a}e^\mathfrak{b}e^\mathfrak{c}e^\mathfrak{d}\right]\;,\label{ym2}
\end{equation}
where $R^\mathfrak{a}_{\phantom{a}\mathfrak{b}}=\mathrm{d}\omega^\mathfrak{a}_{\phantom{a}\mathfrak{b}}+\omega^\mathfrak{a}_{\phantom{a}\mathfrak{c}} \omega^\mathfrak{c}_{\phantom{c}\mathfrak{b}}$ and $T^\mathfrak{a}=\mathrm{d}e^\mathfrak{a}-\omega^\mathfrak{a}_{\phantom{a}\mathfrak{b}}e^\mathfrak{b}$ are the curvature and torsion, respectively. The action \eqref{ym2} can be identified with a four-dimensional gravity, with $\gamma^2=\kappa^2/2\pi G$, where $G$ is the Newton constant. Thus,
\begin{equation}
S=\frac{1}{8\pi G}\int\left[\frac{1}{2\Lambda^2}R^\mathfrak{a}_{\phantom{a}\mathfrak{b}}\star R_\mathfrak{a}^{\phantom{a}\mathfrak{b}}+T^\mathfrak{a}\star T_\mathfrak{a}-\frac{\epsilon}{2}\epsilon_\mathfrak{abcd}R^\mathfrak{ab}e^\mathfrak{c}e^\mathfrak{d}+\frac{\Lambda^2}{4}\epsilon_\mathfrak{abcd}e^\mathfrak{a}e^\mathfrak{b}e^\mathfrak{c}e^\mathfrak{d}\right]\;,\label{ym3}
\end{equation}
where $\Lambda^2=\gamma^2/4$ stands for the cosmological constant.

The field equations are straightforwardly computed. For $e$ we have
\begin{equation}
\frac{1}{4\Lambda^2}R^\mathfrak{bc}\star(R_\mathfrak{bc}e_\mathfrak{a})+\frac{1}{2}T^\mathfrak{b}\star(T_\mathfrak{b}e_\mathfrak{a})+\mathrm{D}\star T_\mathfrak{a}+\epsilon_\mathfrak{abcd}\left(-\epsilon R^\mathfrak{bc}e^\mathfrak{d}+\Lambda^2e^\mathfrak{b}e^\mathfrak{c}e^\mathfrak{d}\right)=0\;,\label{eqe}
\end{equation}
and for $\omega$
\begin{equation}
\frac{1}{2\Lambda^2}\mathrm{D}\star R^\mathfrak{ab}+e^\mathfrak{b}\star T^\mathfrak{a}-\epsilon\epsilon^\mathfrak{ab}_{\phantom{ab}\mathfrak{cd}}T^\mathfrak{c}e^\mathfrak{d}=0\;.\label{eqw}
\end{equation}
From the non-linear character of Eqs.~\eqref{eqe} and \eqref{eqw}, it is evident that, even in the absence of matter, a non-trivial curvature and torsion can be generated. On the other hand, the simplest non-trivial geometry, with $T^\mathfrak{a}=0$, is a constant curvature solution $R^\mathfrak{ab}=2\epsilon\Lambda^2e^\mathfrak{a}e^\mathfrak{b}$. This solution establishes the effective geometry as a Riemannian one. It is widely known that, in the first order formalism, matter can behave as source for torsion. Thus, this solution is in agreement with the absence of matter fields.

\section{Geometrical aspects}\label{formal}

Having covered the basics of the mapping that enables an effective gravity to emerge based on de Sitter gauge theories, we will now briefly discuss some formal aspects of the mechanism previously described.

\subsection{Fiber bundles}\label{fbundles}

We will not provide a review of fiber bundle theory. For that we refer the reader to the available literature \cite{Daniel:1979ez,Kobayashi,Nakahara:1990th}.

de Sitter gauge theories in four Euclidean dimensions are mathematically described as a principal bundle \cite{Daniel:1979ez} $SO=(SO(m,n),\mathbb{R}^4)$ where $SO(m,n)$ characterizes the fiber and structure group while $\mathbb{R}^4$ is the base space identified with spacetime. It is assumed that $SO$ is endowed with a connection 1-form $Y$, recognized as the gauge field.

Gravity, on the other hand, can be defined as a coframe bundle \cite{Trautman:1981fd,Nakahara:1990th}, $C=(GL(4,\mathbb{R}),\mathbb{M}^4)$, where $\mathbb{M}^4$ is a four-dimensional spacetime manifold. The structure group and fiber have a deeper meaning: At each point $X\in\mathbb{M}^4$ one can define the cotangent space $T^*_X(\mathbb{M})$. The fiber is the collection of all coframes $e$ that can be defined in $T^*_X(\mathbb{M})$. These coframes are related to each other through the action of the general linear group. As a consequence, the fiber is actually the group $GL(4,\mathbb{R})$. Accordingly, the coframe bundle is a gauge bundle endowed with a connection 1-form $\Gamma$. Nevertheless, if we look at the fiber as the field of coframes, the coframe bundle is a fiber bundle associated to a tangent bundle \cite{Nakahara:1990th,Kobayashi}. Moreover \cite{Nakahara:1990th}, the action of the structure group from the right are associated with local gauge transformations in $T^*_X(\mathbb{M})$ while the action from the left with general coordinate transformations for $GL(4,\mathbb{R})\subset \mathrm{diff}(4)$. Remarkably, the coframe bundle has a contractible piece $GL(4,\mathbb{R})/SO(4-n,n)$ where $SO(4-n,n)$ is obviously a stability group. This means that the coframe bundle can be naturally contracted down to the orthogonal coframe bundle $C_o=(SO(4-n,n),\mathbb{M}^4)$ \cite{Sobreiro:2010ji,Kobayashi,Nash:1983cq,McInnes:1984kz}.

The starting theory is a de Sitter gauge theory which was reduced by an In\"on\"u-Wigner contraction to $SO^\prime=(SO(m!-1,n),\mathbb{R}^4)$. It immediately follows that the connection $Y$ defined in $SO$ imposes a connection $A$ on $SO^\prime=(SO(m!-1,n),\mathbb{R}^4)$, see \cite{Sobreiro:2010ji,Kobayashi,McInnes:1984kz}. This result can be easily proven from the stability of $SO(m!-1,n)=SO(m,n)/K$, where $K$ is a symmetric coset space that defines an associated bundle $K_{SO}=(SO(m!-1,n),\mathbb{R}^4,K)\equiv SO^\prime\times K$. As a consequence, the field $\theta$ is a section over $K_{SO}$ and thus a matter field coupled to the gauge theory defined in $SO^\prime$, a fact that is evident from transformations \eqref{gt3}.

The induced gravity is obtained from the structures $SO^\prime$, $K_{SO}$ and $\theta$ that can define an orthogonal coframe bundle $C_o$ if and only if the base spaces $\mathbb{R}^4$ and $\mathbb{M}^4$ are isomorphic. The proof is performed by identifying the field $\theta$ at each point $x\in\mathbb{R}^4$ with a coframe field $e$. Thus, the non-triviality of $\theta$ implies a deformation of spacetime $\mathbb{R}^4\longmapsto\mathbb{M}^4$ in such a way that, at each point $X\in\mathbb{M}^4$, the tangent space $T_X(\mathbb{M})$ acquires a local isometry characterized by the gauge group $SO(m!-1,n)$. Thus, the fiber and structure group of the orthogonal coframe bundle are determined by the the field $\theta$ and the gauge group, respectively. In addition, the base space $\mathbb{M}^4$, defined by the action of the vierbein on the tangent space $e:T_X(\mathbb{M})\longmapsto \mathbb{M}^4$, can only exist if, and only if, the mapping $\phi:\mathbb{R}^4\longmapsto\mathbb{M}^4$ is an isomorphism. Otherwise, there will be ambiguities between fibers on the mapping.

\subsection{Uniqueness}

It is crucial to discuss the uniqueness of the effective geometry. The first point concerns the mapping \eqref{id2} and establishes that for each gauge configuration $(A,\theta)$ there will be a geometric configuration $(\omega,e)$. However, it should be emphasized that each configuration $(A,\theta)$ will contribute to the path integral to determine the final quantum action associated to the Yang-Mills action, for instance, as perturbative series,
\begin{equation}
\Gamma=\sum_{n=0}^{\infty}\hbar^n\Gamma^{(n)}\;,\label{qact}
\end{equation}
where the zeroth order coincides with the classical action including the gauge fixing term,
\begin{equation}
\Gamma^{(0)}=S_{\mathrm{YM}}+S_{gf}\;.\label{cact}
\end{equation}
The remnant terms of the series will be responsible for generate the mass parameter $\gamma$ or, eventually, the Gribov problem treatment can be considered in \eqref{cact}. Moreover, the main contribution comes from the Yang-Mills classical action \eqref{ym0} and that is what we have considered in the mapping of Section \ref{induced}. Thus, all gauge configurations and all quantum effects contribute to generate the effective final geometry, which is determined through the field equations \eqref{eqe} and \eqref{eqw}. As a consequence, any solution to these equations can define a different geometry and then an ambiguity might arise. However, that is not the case because the final theory \eqref{ym3} is a classical field theory with its own dynamics, as gravity should be. Thus, these ambiguities must be absent.

Another way to see the absence of ambiguities is the formal analysis developed in Section \ref{fbundles} where the mapping $\mathbb{R}^4\longmapsto\mathbb{M}^4$  is required to be an isomorphism. This requirement avoids two different fibers to overlap. As a consequence, there will be no starting configuration $(A,\theta)$ that generates two sets of geometry.

Finally, we will compute explicitly the mapping ${R}^4\longmapsto\mathbb{M}^4$ and show that it is non-degenerate, as it should be by definition. We have considered that a $p$-form in $\mathbb{R}^4$ is mapped into a $p$-form in $\mathbb{M}^4$ and that their corresponding Hodge duals are identified, as illustrated by \eqref{id2} and \eqref{id4}. Formally, the action \eqref{ym3} is achieved by a map that identifies the space of $p$-forms in $\mathbb{R}^4$, namely $\Pi^p$, into the space of $p$-forms in $\mathbb{M}^4$, $\tilde{\Pi}^p$. In the same way, the Hodge dual space obeys the same map:
\begin{eqnarray}
\Pi^p&\longmapsto&\tilde{\Pi}^p\;,\nonumber\\
*\Pi^p&\longmapsto&\star\tilde{\Pi}^p\;,\label{mapfull}
\end{eqnarray}
For generality purposes, we assume a generic original metric $g_{\mu\nu}$ which, eventually, we can select as a Euclidean metric, and the effective metric is $\tilde{g}_{\mu\nu}$. Moreover, we can also consider manifolds with an arbitrary dimension $d$. Clearly, a necessary extra condition is that both $g=|\det{g_{\mu\nu}}|$ and $\tilde{g}=|\det{\tilde{g}_{\mu\nu}}|$ are non-vanishing quantities.

To find the explicit mapping, we apply the first of \eqref{mapfull} to a generic $p$-form, 
\begin{equation}
f_{\mu_1\ldots\mu_p}(x)dx^{\mu_1}\ldots dx^{\mu_p}=\tilde{f}_{\mu_1\ldots\mu_p}(X)dX^{\mu_1}\ldots dX^{\mu_p}\;,\label{eq1}
\end{equation}
from which one easily obtain
\begin{equation}
f_{\mu_1\ldots\mu_p}(x)=L_{\phantom{\nu_1}\mu_1}^{\nu_1}\ldots L_{\phantom{\nu_p}\mu_p}^{\nu_p}\tilde{f}_{\nu_1\ldots\nu_p}(X)\;,\label{eq3}
\end{equation}
where $L_{\phantom{\nu}\mu}^\nu=\frac{\partial X^\nu}{\partial x^\mu}$. For the corresponding Hodge dual we have,
\begin{eqnarray}
& &\sqrt{g}\epsilon_{\mu_1\ldots\mu_p\nu_{p+1}\ldots\nu_d}f^{\mu_1\ldots\mu_p}(x)dx^{\nu_{p+1}}\ldots dx^{\nu_d}=\nonumber\\
&=&\sqrt{\tilde{g}}\epsilon_{\mu_1\ldots\mu_p\nu_{p+1}\ldots\nu_d}\tilde{f}^{\mu_1\ldots\mu_p}(X)dX^{\nu_{p+1}}\ldots dX^{\nu_d}\;,\label{eq4}
\end{eqnarray}
from which it can be found that
\begin{equation}
f^{\mu_1\ldots\mu_p}=\left({\frac{\tilde{g}}{g}}\right)^{1/2}\left(\frac{L}{d}\right)^{d-p}\tilde{f}^{\mu_1\ldots\mu_p}\;,\label{eq8}
\end{equation}
with $L=L^\mu_{\phantom{\mu}\mu}$.

To compare \eqref{eq3} and \eqref{eq8} we multiply both sides of \eqref{eq8} by the original metric tensor which allows to low the \emph{lhs} indices. To low an index at the \emph{rhs} we extract an effective metric tensor,
\begin{equation}
f_{\mu_1\ldots\mu_p}=\left({\frac{\tilde{g}}{g}}\right)^{1/2}\left(\frac{L}{d}\right)^{d-p}\tilde{g}^{\nu_1\alpha_1}g_{\alpha_1\mu_1}\ldots\tilde{g}^{\nu_p\alpha_p}g_{\alpha_p\mu_p}\tilde{f}_{\nu_1\ldots\nu_p}\;.\label{eq10}
\end{equation}
Combining \eqref{eq10} and \eqref{eq3} we achieve
\begin{equation}
L_{\phantom{\nu}\mu}^\nu=\left({\frac{\tilde{g}}{g}}\right)^{1/2p}\left(\frac{L}{d}\right)^{(d-p)/p}\tilde{g}^{\nu\alpha}g_{\alpha\mu}\;,\label{eq11}
\end{equation}
which is not valid for $p=0$. In that case it easy to find that \eqref{mapfull} is valid only if
\begin{equation}
\left({\frac{\tilde{g}}{g}}\right)^{1/2}\left(\frac{L}{d}\right)^d=1\;.\label{0f}
\end{equation}
The constraint \eqref{0f} implies that
\begin{equation}
L_{\phantom{\nu}\mu}^\nu=\frac{d}{L}\;\tilde{g}^{\nu\alpha}g_{\alpha\mu}\;.\label{eq11a}
\end{equation}
The trace $L$ can now be calculated from \eqref{0f} or \eqref{eq11a} providing
\begin{eqnarray}
L&=&d\left({\frac{g}{\tilde{g}}}\right)^{1/2d}\;,\nonumber\\
L&=&d^{1/2}(\tilde{g}^{\mu\nu}g_{\mu\nu})^{1/2}\;,\label{trace}
\end{eqnarray}
respectively. The relations \eqref{trace} enforce the extra constraint
\begin{equation}
(\tilde{g}^{\mu\nu}g_{\mu\nu})^{1/2}=d^{1/2}\left({\frac{g}{\tilde{g}}}\right)^{1/2d}\;.\label{constx}
\end{equation}
As a consequence, we obtain the final expression for the transformation
\begin{equation}
L_{\phantom{\nu}\mu}^\nu=\left(\frac{\tilde{g}}{g}\right)^{1/2d}\tilde{g}^{\nu\alpha}g_{\alpha\mu}\;.\label{eq13}
\end{equation}
We recall that the effective metric is computed from the field equations, while the original metric is a given quantity\footnote{For the present case we actually have ($d=4$ and $g_{\mu\nu}=\delta_{\mu\nu}$)
\begin{equation}
L_{\phantom{\nu}\mu}^\nu=\left(\tilde{g}\right)^{1/8}\tilde{g}^{\nu\alpha}\delta_{\alpha\mu}\;.\label{eq14}
\end{equation}}. 
It turns out that the mapping \eqref{eq13} has an inverse,
\begin{equation}
\left(L^{-1}\right)_{\phantom{\nu}\mu}^\nu=\left(\frac{g}{\tilde{g}}\right)^{1/2d}g^{\nu\alpha}\tilde{g}_{\alpha\mu}\;.\label{eq15}
\end{equation}
The existence of the inverse ensures the non-degeneracy of the mapping.

\section{Discussion}

We started with a standard gauge theory in a Euclidean spacetime. In that situation, the theory is actually renormalizable, at least through all orders in perturbation theory. As a non-Abelian gauge theory, it presents asymptotic freedom and the possibility of dynamical mass generation. Then, a proposition for quantum gravity has been made as long as it induces an effective geometry that could be interpreted as gravity. The fact that the theory possesses a mass parameter enabled the rescaling \eqref{resc1} that, together with asymptotic freedom, allows the deformation of the de Sitter algebra at low energies. This deformation actually induces a symmetry breaking for the Lorentz group, which finally allows the identification of the fundamental fields with geometric quantities \eqref{id2}. Thus, it was formally shown that the dynamical content of a pure gauge theory can induce an effective geometry.

To show that this geometry is actually a gravity theory is a mathematical exercise of fiber bundle theory. The formal analysis is displayed in Section \ref{formal}. At each point $x\in\mathbb{R}^4$ it is defined, through \eqref{id2}, a vierbein and spin connection, which induces a deformation of the spacetime. That is, the spacetime is deformed to a generic differential manifold $\mathbb{M}^4$ with coordinates $X$. At $X\in\mathbb{M}^4$, a tangent space $T_X(\mathbb{M})$ is defined through its local isometries that are characterized by the broken gauge group $SO(m!-1,n)$. The actual geometry is then determined from the field equations \eqref{eqe} and \eqref{eqw}.

The effective gravity theory is described by the action \eqref{ym3} whose simplest nontrivial solution is (anti) de Sitter spacetime, depending on the values of $\Lambda^2$ and $m$. The fact that the gauge group determines the local isometries has a remarkable consequence: For the cases $m\in\{0,1\}$ the reduced group is $SO(4)$ implying that the local isometries are that of an Euclidean space. On the other hand, for $m=2$, it is the Lorentz group $SO(1,3)$ that determines the local isometries. As a consequence of the latter case, space and time are then explicitly distinct from each other. Thus, we started from a gauge theory in a space where space and time are totally mixed (In the sense that there is no physical effect capable of distinguish them from each other), and end up with a theory where a local Minkowski metric appears naturally. This effect can be interpreted as the rising of the equivalence principle where gravity deforms spacetime even if inertial reference frames can be defined locally, \emph{i.e.}, a deformed spacetime which is locally Minkowskian flat. 

We note a recent work, by E.~W.~Mielke \cite{Mielke:2011zz}, where another mechanism in deriving an effective gravity in four dimensions was developed. In this work, the gauge group is the $SL(5,\mathbb{R})$ and the action is, essentially, the corresponding BF action \cite{q-alg/9507006,NSF-ITP-88-178} complemented with the Pontrjagin density and an algebra-valued Higgs field. Besides the gauge group choice and the starting action, this model differs from ours also by the fundamental mechanism that makes gravity to emerge. In the case of \cite{Mielke:2011zz}, gravity is obtained from a Higgs mechanism, while in the present case it is a dynamical mechanism that generates geometry. It is not our intent to decide which theory is better. It is our opinion, however, that each technique has its qualities. For instance, Mielke's starting action is topological, and the fact that it generates gravity is really astounding. However, this topological action has already a mass scale $1/\mathfrak{\ell}$, which is a necessary condition for a Higgs mechanism to occur. The final result is a gravity theory with extra matter fields. In particular, the Higgs mechanism is necessary in order to break the topological character of the theory. Otherwise, gravity would never be generated. In our case, the starting action is a massless Yang-Mills action with no topological character \eqref{ym0}. The dynamical content of the theory is that generates a possible gravity theory with extra terms, but no extra matter fields. The important step is that a mass parameter emerges dynamically, which makes possible the mapping to a gravity theory.

One of the most important features of $SU(N)$ gauge theories is unitarity \cite{Itzykson:1980rh}, a property that, among other features, follows from the compact character of the gauge group. In the case of the present theory, unitarity is only ensured for $m=0$. In that case, the resulting gravity theory is an $SO(4)$ local isometric gravity. The local Euclidean character of spacetime provides a kind of incomplete prediction of the equivalence principle because it lacks the split between space and time\footnote{The case $m=1$ provides a non-unitary theory and also a local Euclidean spacetime with no difference between space and time.}. On the other hand, if we start with $m=2$, the resulting theory is that of $SO(1,3)$ local isometries. That is, the local spacetime is a Minkowski space and thus the split is a direct consequence that rises together with the equivalence principle. Thus, if on the one hand, we start with a perfect quantum theory, the resulting gravity is not exactly the desired one. On the other hand, by giving up unitarity, the split between space and time correctly emerges.

The case $m=0$ is obviously the most convenient for a starting theory, and the case $m=2$ is the desirable final theory. The technique that would connect both endings is a Wick rotation to an imaginary time, which has no evident physical justification. Moreover, making the rotation before the mapping, at quantum level, is entirely different from the rotation after the mapping. In the first case, the rotation is performed in the gauge group, and there is no relation with time. Then, the afterwards-mapped gravitational time arises with the expected sign. However, such rotation would spoil unitarity in quantum computations. In the second case, the rotation is performed directly in the time coordinate, and the problem that rises is that of dealing with an imaginary time. The apparent paradox can be solved if one imposes that the rotation is performed during the mapping $\mathbb{R}^4\longmapsto\mathbb{M}^4$. In that case, unitarity and quantum computations are consistent, while the resulting gravity theory produces a complete emergent equivalence principle. Thus, the local tangent isometries are determined by the relation $SO(m!-1,n)\longmapsto SO(1,3)$ which consists on a map from a 4-sphere into a 4-hyperbola for $m\in\{0,1\}$ and a hyperbola identity map for $m=2$.

It is also remarkable that Newton's and cosmological constants can be actually computed from the standard quantum field theory techniques, at least at perturbative level. In the case of Newton's constant, $G=\kappa^2/2\pi\gamma^2$, a consistent value can be obtained for the perturbative sector of the theory while for the cosmological constant a bound is obtained, $\lambda^2=\kappa^2/8\pi G$. Thus, if a small value for $G$ is found, then $\Lambda^2$ is large and might compensate for the quantum field theory predictions in order to generate an effective cosmological constant consistent with astrophysical observations \cite{hep-th/0603057,arXiv:1004.1493,Ma:2007pe,Perivolaropoulos:2008pg,arXiv:1001.4538,Luongo:2011yk,Luongo:2011yk2}. In particular, an interesting method to compute the renormalized cosmological constant would be that one discussed in \cite{gr-qc/0611055,Shapiro:2009dh}, where the renormalized cosmological constant should be a compensation between the observational and the quantum field theory cosmological constants, namely $\Lambda_{ren}=\Lambda_{obs}-\Lambda_{qft}$. However, we have assumed that the ratio $\gamma^2/\kappa^2$ is small at low energies, which leads to a large $G$. Notwithstanding, it is really difficult to determine the true behavior of $\kappa$ at non-perturbative level. If it behaves as the strong coupling parameter from $SU(N)$ gauge theories \cite{Cucchieri:2009zt}, it follows that there will be a certain region that $\kappa$ is actually big and the In\"on\"u-Wigner contraction can be performed. Beyond that scale, the coupling actually diminishes until it reaches a finite value at the origin. Thus, after the contraction, there is a chance that $G$ may be small. Obviously, the mass has also a crucial role in the determination of $G$. Moreover, as a quantum gravity prototype, many scales are expected. That is, due to the intensity of gravitational force, it is expected that its classical limit is attained before Planck scale. Thus, the present behavior of gravity should be at the deep infrared sector of the theory. Then, a complete non-perturbative treatment may be required. Moreover, we emphasize the need for further studies for matching cosmological data with our theoretical results. We postpone to a forthcoming work the determination of the physical origin of $\gamma$ and the perturbative loop expansion (and perhaps a semi-perturbative treatment)  in order to find the possible solutions for $\kappa$ and $\gamma$. For the time being, we can only say that the present theory is stable at the quantum level and can actually generate an effective gravity theory.

\section*{Acknowledgements}

The authors are in debt to Prof.~Yu.~Obukhov for clarifying some points in reference \cite{Obukhov:1998gx}. We are grateful to Prof.~ H.~S.~Smith and J.~A.~Helay\"el-Neto for the text review. Conselho Nacional de Desenvolvimento Cient\'{i}fico e Tecnol\'{o}gico\footnote{RFS is a level PQ-2 researcher under the program \emph{Produtividade em Pesquisa}, 304924/2009-1.} (CNPq-Brazil) and the Coordena\c{c}\~ao de Aperfei\c{c}oamento de Pessoal de N\'{\i}vel Superior (CAPES) are acknowledged for financial support. RFS is also partially supported by Pro-Reitoria de Pesquisa, P\'os-Gradua\c{c}\~ao e Inova\c{c}\~ao\footnote{ Under the program \emph{Jovens Pesquisadores 2009}, project 305.} of the Universidade Federal Fluminense (Proppi-UFF).

\end{document}